\documentclass[aps,prb,twocolumn,superscriptaddress,longbibliography]{revtex4-2}
\usepackage[utf8]{inputenc}
\usepackage{epsfig,amsopn}
\usepackage{graphicx}
\usepackage{physics}
\usepackage{color}
\usepackage{amsmath,amssymb}
\usepackage{enumerate}
\newcommand\bea{\begin{eqnarray}}
\newcommand\eea{\end{eqnarray}}
\newcommand\beq{\begin{equation}}
\usepackage{gensymb}
\usepackage{verbatim}
\newcommand\eeq{\end{equation}}

\def\nn{\nonumber}

\def\si{\sigma}
\def\Do{\partial}

\def\ua{\uparrow}
\def\da{\downarrow}

\begin{document}
\title{Geometry induced net spin polarization of $d$-wave altermagnets}
 \author{ Abhiram Soori}
 \email{abhirams@uohyd.ac.in}
 \affiliation{School of Physics, University of Hyderabad, Prof. C. R. Rao Road, Gachibowli, Hyderabad-500046, India}

\begin{abstract}
Altermagnets exhibit spin-split electronic bandstructures despite having zero net magnetization, making them attractive for field-free spintronic applications. In this work, we show that a finite rectangular altermagnetic sample can acquire a net spin polarization purely due to its geometry. This effect arises from the interplay between the anisotropic, spin-resolved Fermi contours of an altermagnet, the discrete sampling of momentum space and unequal sample dimensions. By explicitly counting occupied states, we demonstrate that rectangular samples with $L_x \neq L_y$ host a finite spin polarization, which vanishes in the symmetric limit $L_x=L_y$ and in the thermodynamic limit. We further show that this geometry-induced spin polarization can be directly probed in transport measurements. In the tunneling regime, the charge and the spin conductances exhibit characteristic   patterns as a function of sample dimensions, faithfully reflecting the underlying spin polarization. In addition, transport across ferromagnet--altermagnet--ferromagnet junctions reveals an asymmetric magnetoresistance with respect to reversal of the Zeeman field, providing an independent transport signature of the finite spin polarization. Our results establish geometry as an effective control parameter for spin polarization in altermagnets and suggest a viable route for exploiting finite-size effects in mesoscopic altermagnetic spintronic devices.
\end{abstract}

\maketitle

\section{Introduction}

Altermagnets (AMs) are a recently identified class of magnetic materials that exhibit spin-split electronic bandstructures despite having zero net magnetization, originating from magnetic order that breaks time-reversal symmetry while preserving combined symmetries~\cite{smejkal22b}. In particular, $d$-wave altermagnets host a $d$-wave magnetic order, drawing a close analogy to unconventional superconductors with $d$-wave pairing symmetry. The absence of net magnetization is a defining characteristic of AMs and renders them especially attractive for field-free spintronic applications, where stray magnetic fields are undesirable~\cite{qfsun2024,nowak2025,sahoo2025tjde,nag2025}.

The spin-split nature of the electronic structure in AMs has led to a growing interest in their transport properties and device potential. Notably, it enables the generation of spin currents in hybrid junctions involving normal metals, even though both constituent materials individually possess zero net spin polarization~\cite{das2023}. Furthermore, the spin-dependent band anisotropy gives rise to orientation-dependent Andreev reflection at AM--superconductor interfaces~\cite{papaj23,sun23} and can strongly enhance crossed Andreev reflection in altermagnet-based heterostructures by suitably controlling the relative orientations of the magnetic order~\cite{Das24,niu2024car}. Electron transport across AMs has been explored in a variety of settings, both theoretically~\cite{das2023,qfsun2025} and experimentally~\cite{xin2025,noh2025}, establishing AMs as a promising platform for spin-dependent transport phenomena.

Beyond these effects, the anisotropic nature of the spin-resolved bandstructures in altermagnets provides an additional and largely unexplored route to controlling spin polarization. In particular, when the electronic spectrum is sampled discretely, as in finite-sized systems, the anisotropy of the Fermi contours can lead to an unequal population of the two spin species. This opens up the possibility of inducing a net spin polarization purely through sample geometry, without invoking external magnetic fields or intrinsic magnetization. Related finite-size effects have been identified in altermagnetic strips, where a net local magnetization develops near the edges~\cite{linder2024}. In contrast, we focus on a geometry-induced spin polarization that arises from the global shape and aspect ratio of the sample and manifests itself in transport measurements. Specifically, we analyze this effect in finite altermagnetic samples and propose experimentally accessible schemes to detect it, highlighting finite-size and shape effects as effective control parameters in altermagnet-based spintronic devices.

\section{Spin polarization of a two-dimensional rectangular AM}

In this section, we demonstrate that a two-dimensional altermagnet (AM) with a rectangular geometry can acquire a finite net spin polarization. The low-energy Hamiltonian describing the AM is given by
\bea
H &=& -t a^2(\partial_x^2+\partial_y^2)\sigma_0  
     + t_J a^2 (\partial_x^2-\partial_y^2) \sigma_z ,
\eea
where $t>0$ sets the kinetic energy scale, $t_J$ ($0<t_J<t$) characterizes the strength of the altermagnetic order~\cite{smejkal22b} which results in spin-split, anisotropic bandstructure, $a$ is a fundamental length scale corresponding to the lattice spacing, and $\sigma_i$ are Pauli matrices in spin space.

We consider a rectangular two-dimensional AM sample of length $L_x$ and width $L_y$. Imposing periodic boundary conditions along both directions, the allowed wavevectors are $k_x=n_x\pi/L_x$ and $k_y=n_y\pi/L_y$, where $n_x,n_y$ are integers. The dispersion relations for electrons with spin $\sigma$ ($\sigma=\uparrow,\downarrow$) are then given by
\bea
E_\sigma(\vec{k}) &=& t a^2 (k_x^2+k_y^2)  - s_\sigma t_J a^2 (k_x^2-k_y^2),~~
\eea
with $s_\uparrow=+1$ and $s_\downarrow=-1$.

\begin{figure}
\includegraphics[width=8cm]{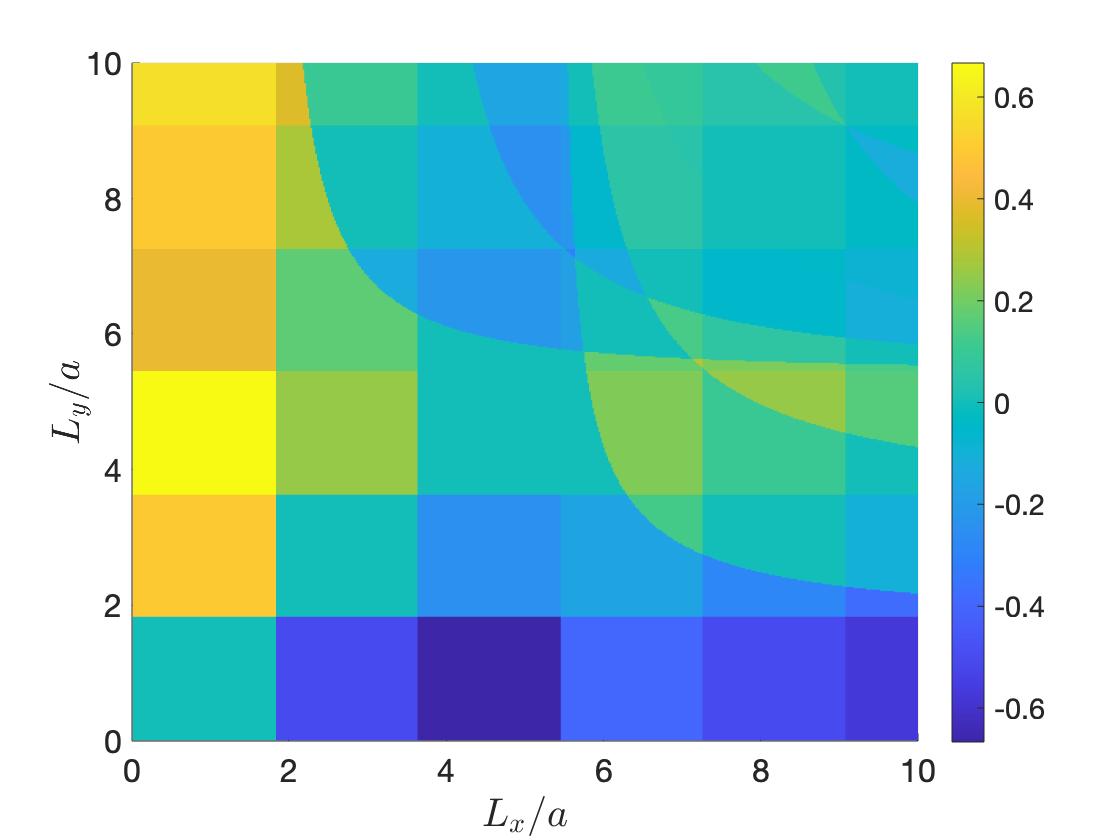}
\caption{Normalized spin polarization $(N_\uparrow-N_\downarrow)/(N_\uparrow+N_\downarrow)$ versus $(L_x,L_y)$ for $t_J=0.75t$ and  $E_F=0.4t$ with periodic boundary conditions. The samples with $L_x\neq L_y$ exhibit nonzero spin polarization. }\label{fig:Ndif}
\end{figure}

For a given Fermi energy $E_F$, we numerically evaluate the total number of occupied states $N_\sigma$ for each spin species by counting the allowed $(k_x,k_y)$ modes satisfying $E_\sigma(k_x,k_y)<E_F$. Figure~\ref{fig:Ndif} shows the normalized spin polarization $(N_\uparrow-N_\downarrow)/(N_\uparrow+N_\downarrow)$ as a function of the sample dimensions $(L_x,L_y)$ for realistic parameter values of $t_J=0.75t$ and $E_F=0.4t$ for KRu$_4$O$_8$~\cite{smejkal22b,weber2024,ghadi2026}.

For $L_x=L_y$, the net spin polarization vanishes due to the geometric symmetry between the two directions. When $L_x \neq L_y$, this symmetry is broken and one direction is preferentially selected. In a rectangular sample, the discretization of momentum space becomes anisotropic, with spacings $\Delta k_x \sim \pi/L_x$ and $\Delta k_y \sim \pi/L_y$. Because the spin-resolved Fermi contours in an altermagnet are anisotropic and elongated along orthogonal directions for opposite spin species, this geometry-induced discretization affects the two spins differently. In particular, for the spin whose Fermi contour is elongated along the $x$ direction, a larger $L_x$ results in a finer sampling of momenta along $k_x$, allowing more states to satisfy $E(\vec k)<E_F$ for a given range of $k_y$. In contrast, the opposite spin, whose Fermi contour is elongated along the $y$ direction, benefits less from this enhanced resolution in $k_x$. Consequently, an imbalance in the number of occupied states of the two spin species develops, leading to a finite net spin polarization.

As a result, the imbalance in the number of occupied states depends sensitively on the aspect ratio of the sample, and the normalized spin polarization increases with $|L_x/L_y - 1|$, vanishing in the symmetric limit $L_x = L_y$. However, it is important to note that the spin species with a Fermi contour elongated along the $y$ direction samples a correspondingly larger range of $k_y$, which compensates for the reduced resolution along $k_x$. Consequently, when the system size becomes large, the total number of states for the two spin species scales similarly, and the net spin polarization per unit area approaches zero. This highlights that the geometry-induced spin polarization is predominantly a finite-size effect, most pronounced in small samples.

\begin{figure}
\includegraphics[width=8cm]{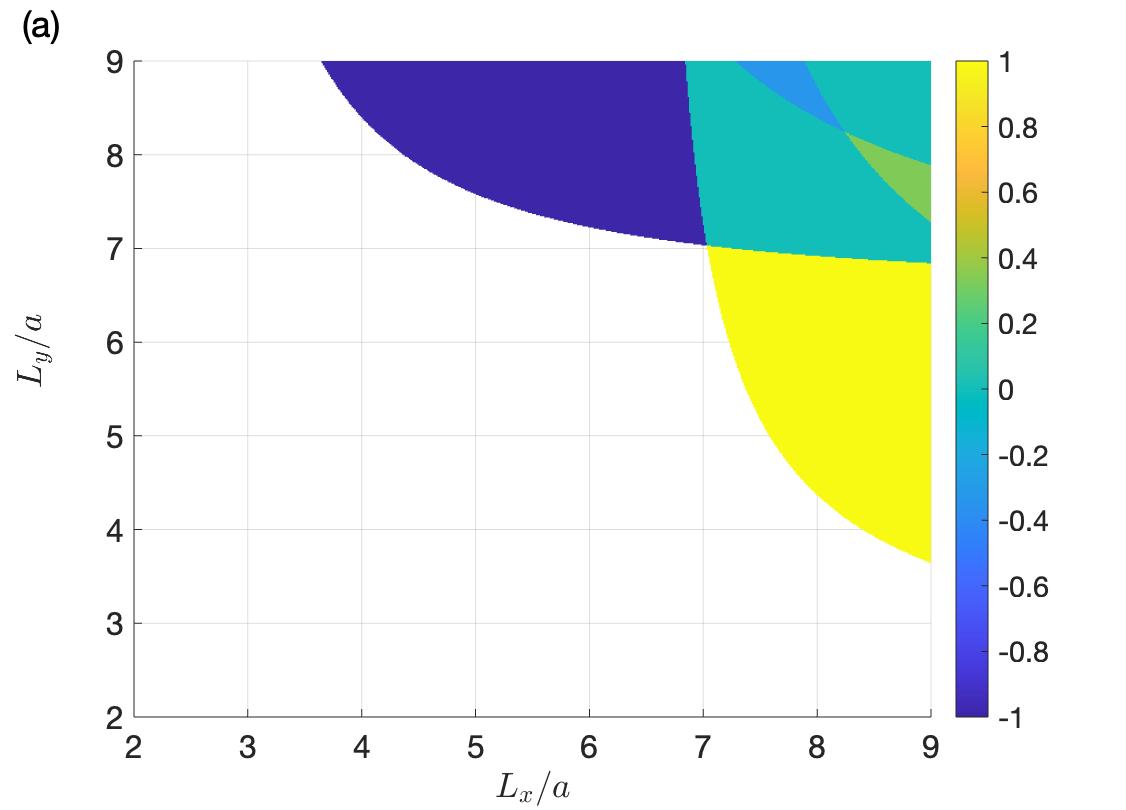}
\includegraphics[width=4cm]{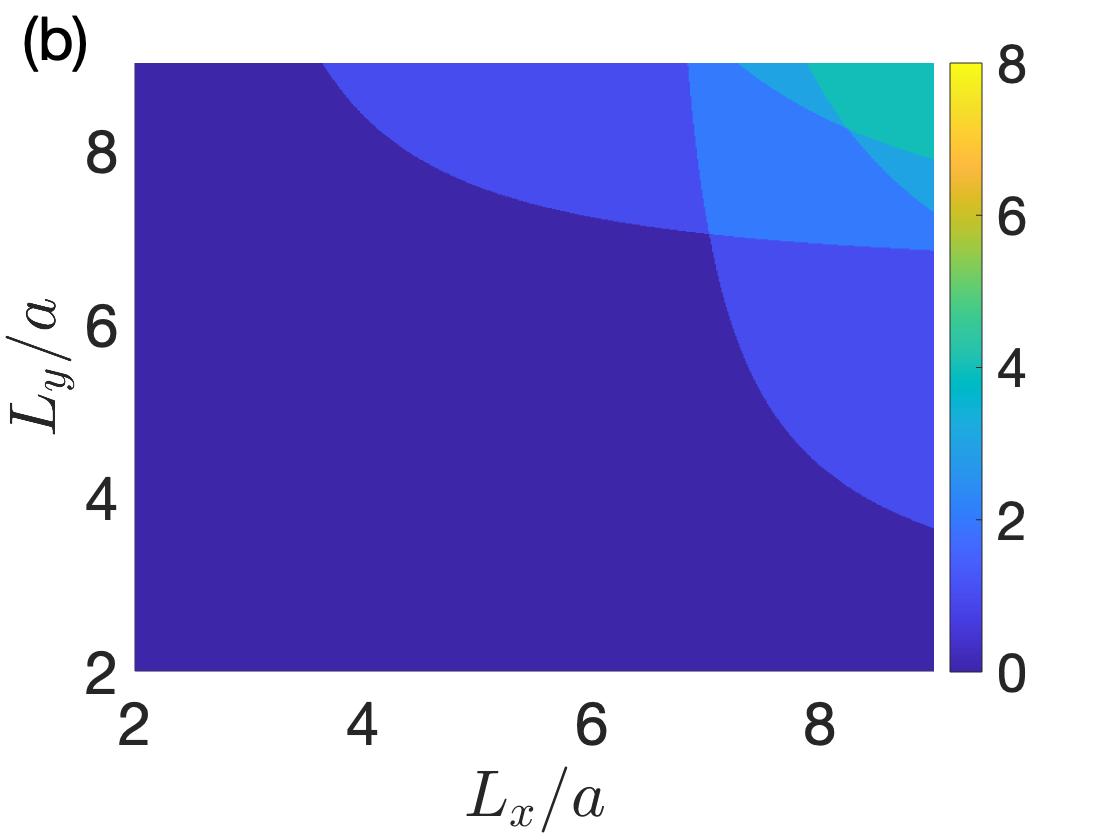}
\includegraphics[width=4cm]{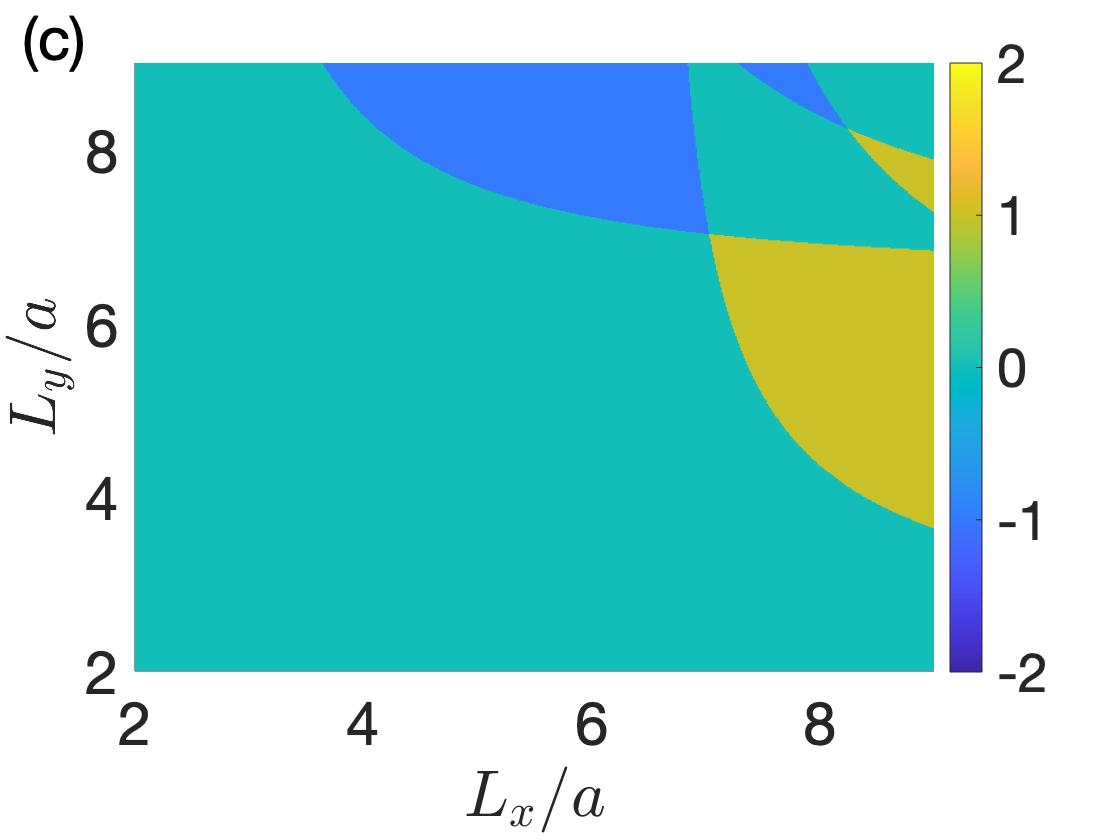}
\caption{(a) Normalized spin polarization $(N_\uparrow-N_\downarrow)/(N_\uparrow+N_\downarrow)$ (b) $(N_{\ua}+N_{\da})$ (c) $(N_{\ua}-N_{\da})$, versus the sample dimensions $(L_x, L_y)$. The samples with $L_x\neq L_y$ exhibit nonzero spin polarization. }\label{fig:Ndif-obc}
\end{figure}

When periodic boundary conditions are imposed, the allowed momenta are quantized as
\(k_j = n_j \pi / L_j\) for \(j = x,y\), where \(n_j\) takes integer values, including zero.
The condition \(E(\vec{k}) < E_F\) then translates into
\(|n_j| < k_{F,j} L_j / \pi\) for modes with \(n_{j'} = 0\), where \(j' = y,x\) for
\(j = x,y\), and the spin-dependent Fermi wavevectors are given by
\(k_{F,x} = \sqrt{E_F / [(t - s_\sigma t_J)a^2]}\) and
\(k_{F,y} = \sqrt{E_F / [(t + s_\sigma t_J)a^2]}\).
Consequently, the square-like features observed in Fig.~\ref{fig:Ndif} originate from
modes with either \(n_x = 0\) or \(n_y = 0\).
This behavior changes qualitatively when open boundary conditions are imposed, for which
\(n_j\) is restricted to positive integers and the \(n_j = 0\) modes are excluded.
As a result, the square-like pattern disappears in the normalized spin polarization plotted
as a function of sample dimensions, as shown in Fig.~\ref{fig:Ndif-obc}. This distinction is 
particularly relevant for experiments, since periodic boundary conditions are an idealization 
and realistic finite samples are governed by open boundary conditions, for which the 
square-like features are absent.

\section{Transport probe of geometry-induced spin polarization}~\label{sec:transport}
\begin{figure}
\includegraphics[width=8cm]{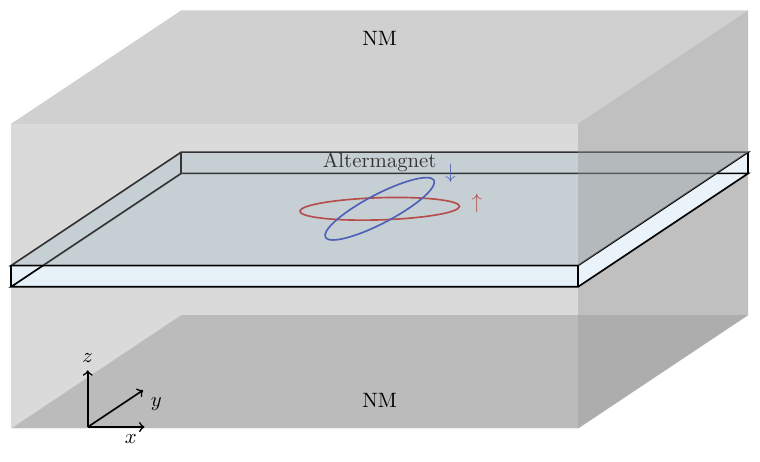}
\caption{A thin rectangular slab of AM connected to NM electrodes with rectangular cross-section of same size. The dimensions of the AM slab is $L_x\times L_y \times d$. The NM leads are extended along $z$-direction and have same cross sectional dimensions as that of AM. }\label{fig:transport}
\end{figure}

An experimental probe of the geometry-induced spin polarization in AMs is essential not only for its direct detection but also for assessing its potential utility in spintronic applications. To this end, we propose a transport-based setup in which a rectangular AM sample is sandwiched between two normal-metal electrodes with rectangular cross sections, as shown in Fig.~\ref{fig:transport}. We demonstrate that electron transport through the AM in this geometry exhibits characteristic signatures that directly reflect the presence of a net spin polarization in the AM.

The Hamiltonian to study transport in such a setup is given by 
\bea 
H &=& 
\begin{cases}
-t_na^2(\Do_x^2+\Do_y^2+\Do_z^2)\si_0, ~~~{\rm for ~~~} |z|>d/2, \\ ~~\\
-t a^2(\partial_x^2+\partial_y^2+\Do_z^2)\sigma_0
      + t_J a^2(\partial_x^2-\partial_y^2)\sigma_z \\~~~~~~~~~~~~~~~~~~~~~~~~~~~~~~~~~{\rm for~~} |z|<d/2, 
\end{cases}\label{eq:H-transport}
\eea
where $d$ is the thickness of the AM slab and $t_n$ is the kinetic energy scale in NM electrodes. The boundary conditions to solve the scattering problem at $z=\pm d/2$ are given by 
\bea
\Do_z\psi|_{z_0^+}-\Do_z\psi|_{z_0^-} &=& q_0\psi(z_0), ~~~~{\rm for} ~~z_0=\pm d/2, \label{eq:bc}
\eea
in addition to the continuity of $\psi$. 
For all $z$, the system is confined to $0<x<L_x$ and $0<y<L_y$. We impose open boundary conditions along the $x$ and $y$ directions, as these provide a more realistic description of a finite sample. Along the $z$ direction, the AM slab is subject to open boundary conditions. As a consequence, the lowest allowed transverse mode acquires a finite energy shift $t a^2 \pi^2/d^2$ compared to the strictly two-dimensional case, where $d$ denotes the thickness of the AM slab.

\begin{figure}
\includegraphics[width=7cm]{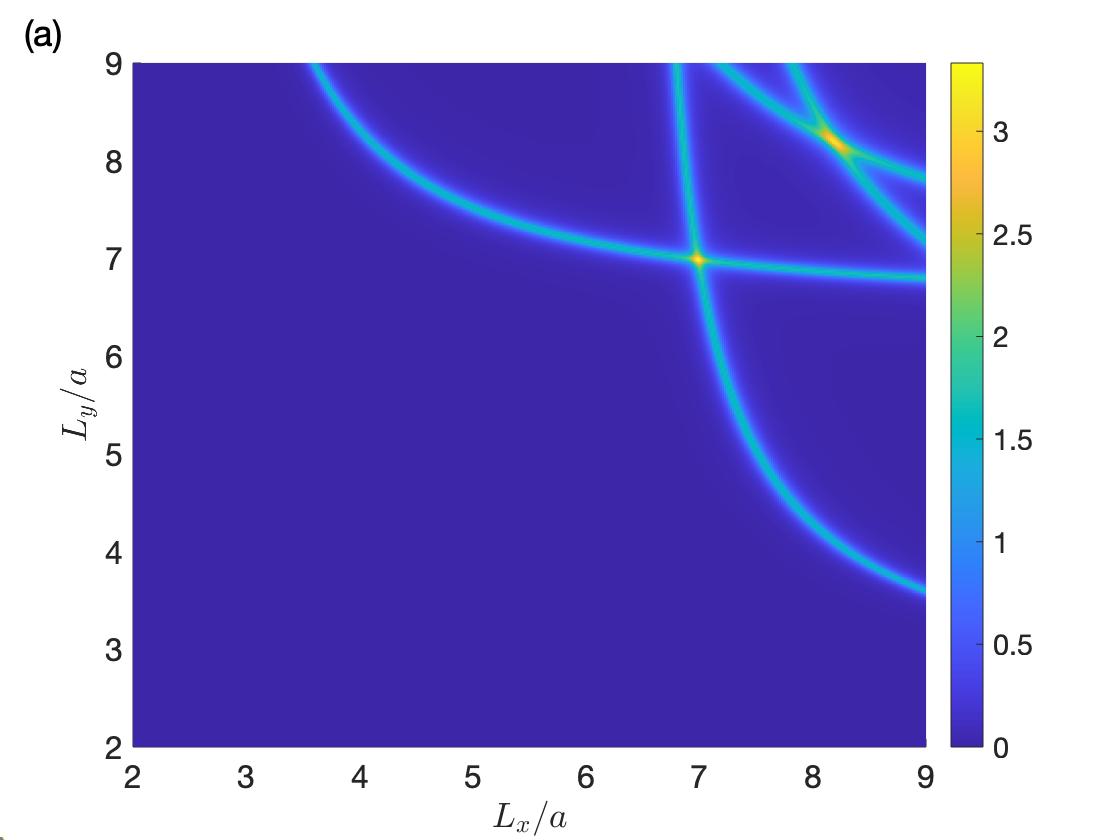}
\includegraphics[width=7cm]{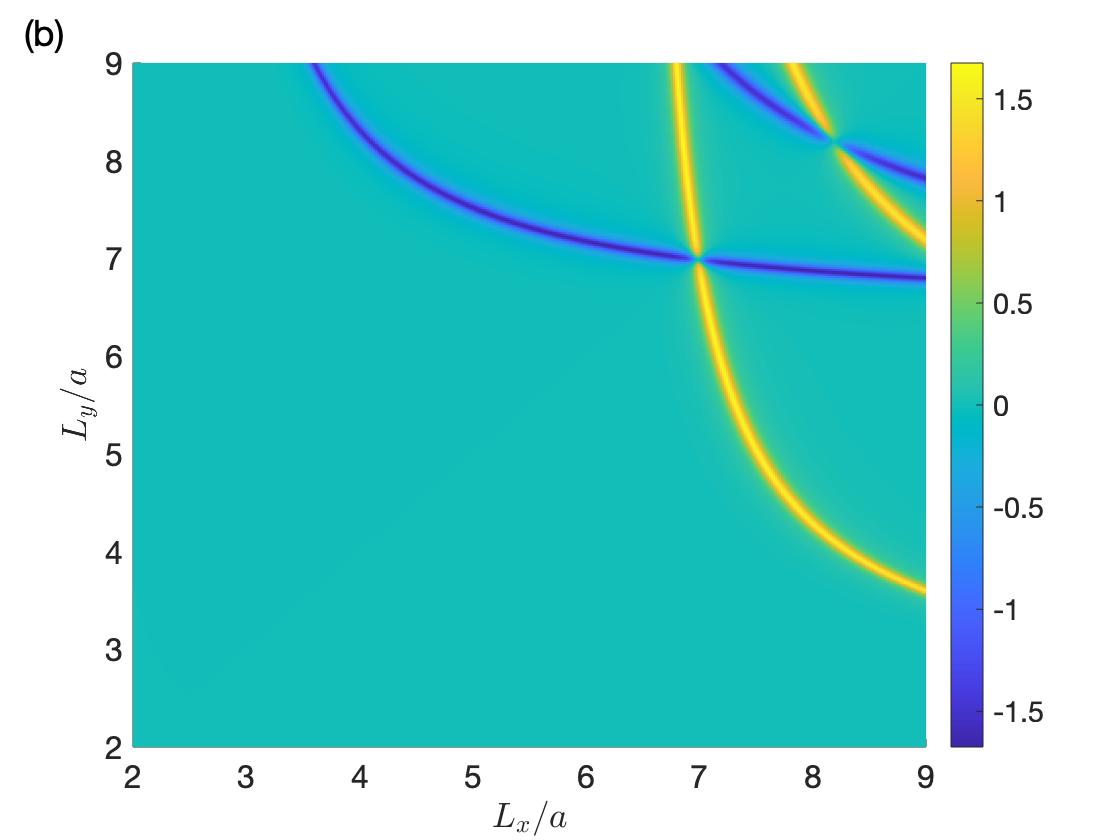}
\caption{ (a) Charge conductance in units of $e^2/h$ and (b) spin conductance in units of $e/2$ across AM versus the system size $(L_x,L_y)$ for $t_n=0.2t$, $t_J=0.75t$, and $d=5a$ for the system with open boundary conditions along $x$ and $y$.  }\label{fig:GsG}
\end{figure} 

The transport properties are evaluated using the Landauer formalism~\cite{landauer,datta}, with the conductance calculated independently in the two spin sectors. The sum of the two contributions defines the charge conductance $G$, while their difference yields the spin conductance $G^s$.  
In Fig.~\ref{fig:GsG}, we plot (a) $G$ and (b) $G_s$ versus $(L_x,L_y)$ for $q_0=10$, $E_F=t$, with $t_n=0.2t$, $t_J=0.75t$, and $d=5a$. The hopping parameter $t_n$ in the normal-metal leads is chosen such that their Fermi surface fully encloses that of the AM, ensuring that all relevant AM bands participate in transport.

As evident from Fig.~\ref{fig:GsG}, both the charge conductance $G$ and the spin conductance $G_s$ exhibit extrema along contours that coincide with the transitions in the number of occupied states shown in Fig.~\ref{fig:Ndif-obc}.  Along these contours, the energy of a state in the isolated AM sample matches the energy of electrons incident from the leads, giving rise to transmission resonances. We emphasize that this direct correspondence between spin polarization and transport signatures emerges only in the tunneling limit. 

Next, we study transport across an FM--AM--FM junction, where the normal-metal electrodes in Fig.~\ref{fig:transport} are replaced by ferromagnetic (FM) leads. The Hamiltonian for the FM electrodes is
\bea
H=-t_n a^2(\partial_x^2+\partial_y^2+\partial_z^2)\sigma_0 - b\,\sigma_z, \label{eq:HFM}
\eea
valid in the regions $|z|>d/2$. Here, $b$ denotes the Zeeman energy scale, proportional to the magnetization of the FM leads. All other details of the setup remain unchanged.

\begin{figure}
\includegraphics[width=4cm]{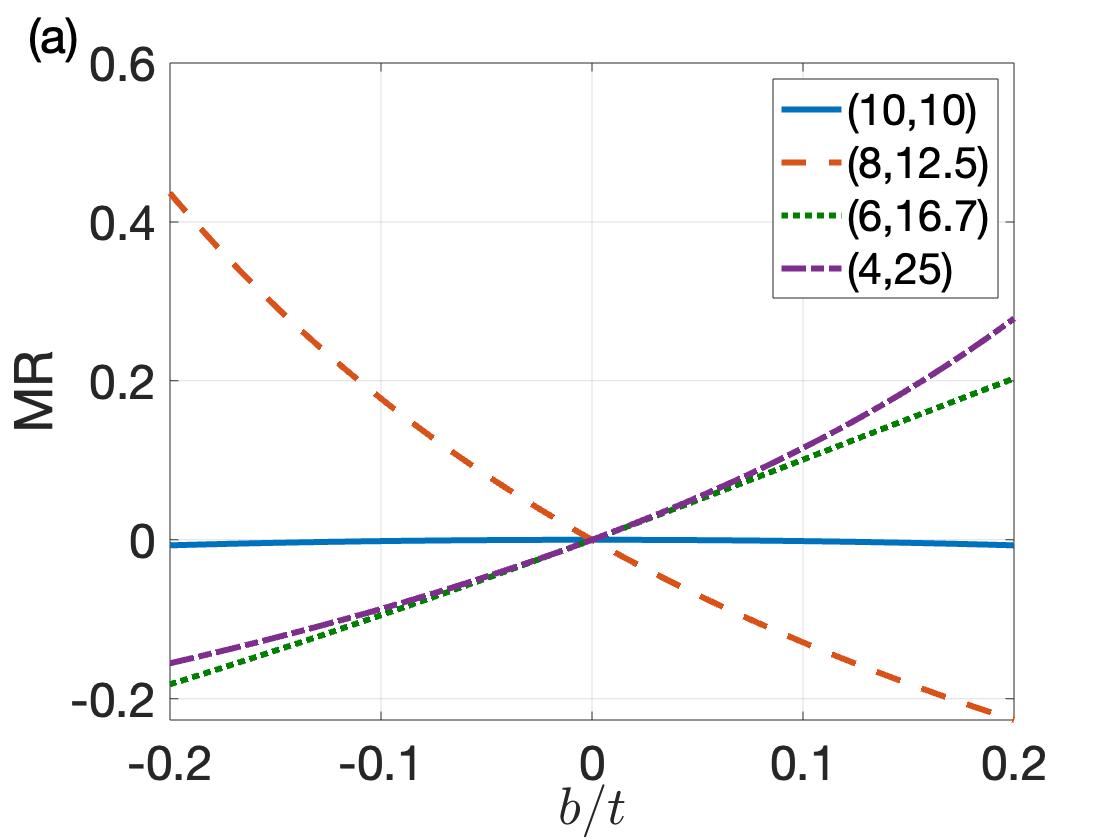}
\includegraphics[width=4cm]{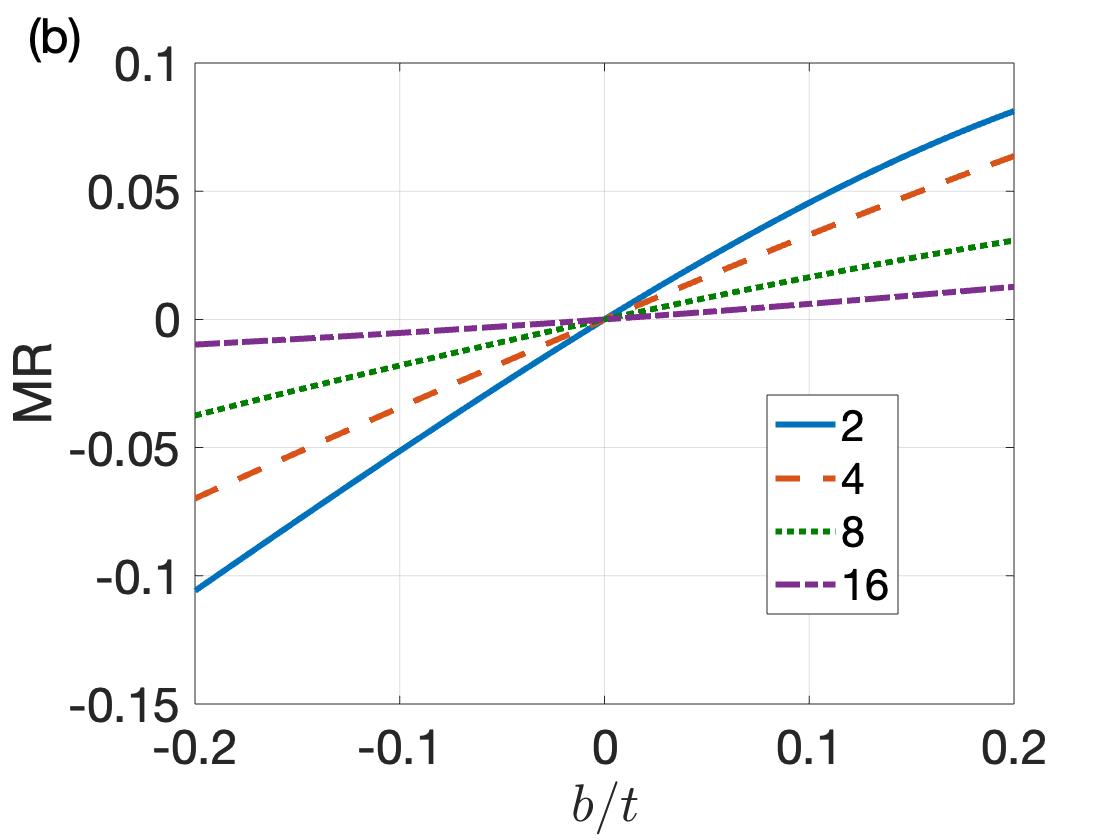}
\includegraphics[width=9cm]{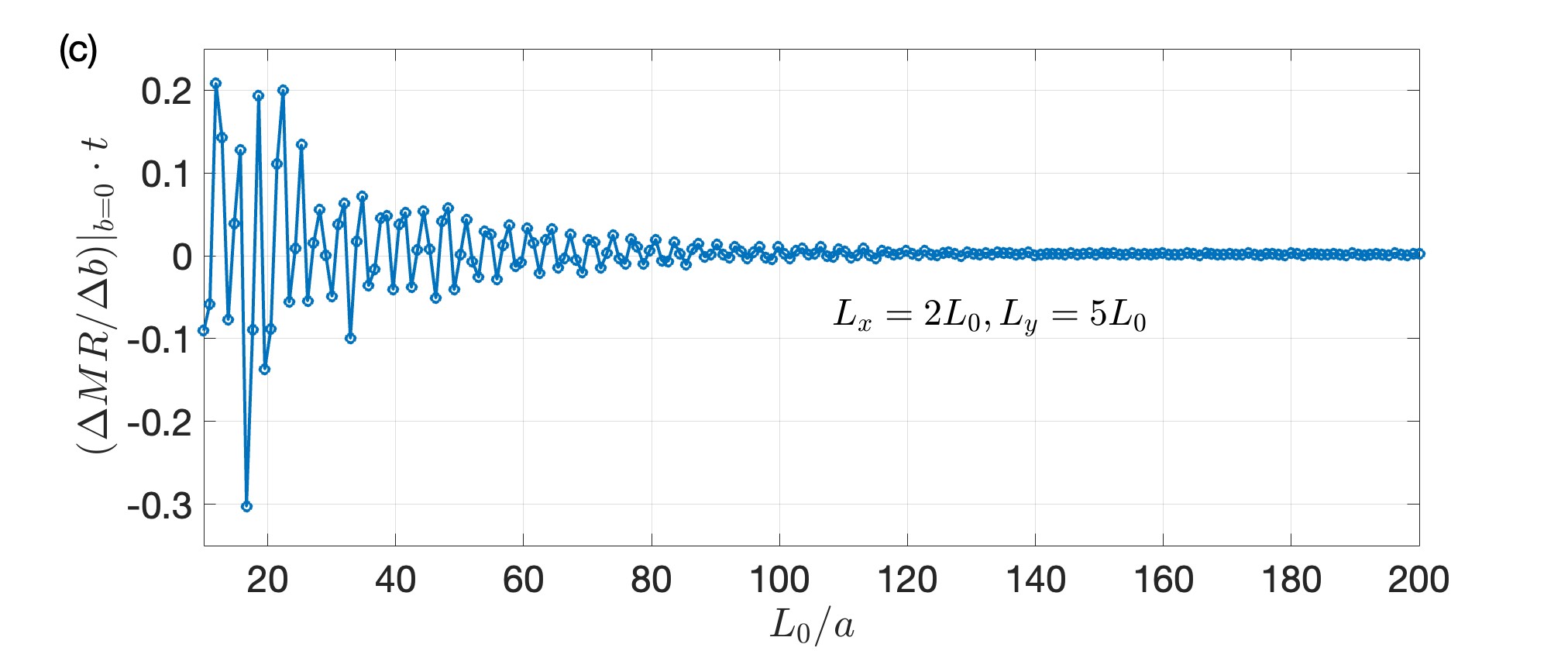}
\caption{
Magnetoresistance (MR) of the FM--AM--FM junction. 
(a) MR as a function of the Zeeman energy $b$ for different system sizes. 
Each curve corresponds to a distinct choice of $(L_x,L_y)/a$, as indicated in the legend. 
(b) MR versus $b$ for systems with the same aspect ratio, $(L_x,L_y)=n(4,10)a$, for different values of $n$ shown in the legend. 
(c) The asymmetry measure $[\,\mathrm{MR}(b)-\mathrm{MR}(-b)\,]/2b$ as a function of the characteristic system size $L_0$, for $b=0.1t$. It can be seen that in the limit of large system sizes the asymmetry measure tends to zero. 
All other parameters are the same as in Fig.~\ref{fig:GsG}.
 }\label{fig:MR}
\end{figure}

In our transport calculations, the FM leads and the altermagnetic region are treated as noninteracting subsystems, and the interface is modeled through the matching of scattering states at the boundaries. We do not include a self-consistent proximity-induced modification of the magnetic order parameters.

We compute the conductance $G(b)$ as a function of the Zeeman energy $b$ and characterize the transport response via the magnetoresistance (MR), defined as
$\mathrm{MR}=\delta R(b)/R(0)$, where $\delta R(b)=R(b)-R(0)$ and $R(b)=1/G(b)$ is the resistance at Zeeman energy $b$. 
In Fig.~\ref{fig:MR}(a), we plot the MR as a function of $b$ for several choices of the aspect ratio $L_x/L_y$, while keeping the cross-sectional area fixed. 
For rectangular samples with $L_x\neq L_y$, the MR is clearly asymmetric under $b\to -b$, reflecting the presence of a nonzero net spin polarization in the altermagnet. 
In contrast, for $L_x=L_y$, the MR is symmetric in $b$, consistent with the absence of net spin polarization.

In Fig.~\ref{fig:MR}(b), we plot the MR as a function of $b$ for several system sizes sharing the same aspect ratio. 
As the system size increases, the slope of the MR curve at $b=0$ is progressively reduced. 
We quantify this asymmetry using the slope at $b=0$, which may be approximated by $[\,\mathrm{MR}(b)-\mathrm{MR}(-b)\,]/2b$ for small $b$. 
Figure~\ref{fig:MR}(c) shows this asymmetry measure as a function of system size for a fixed aspect ratio. 
In the limit of large system sizes, the asymmetry tends to zero, indicating that the net spin polarization in altermagnets is predominantly a finite-size effect.

\section{Spin polarization for different altermagnetic symmetries}

Altermagnets (AMs) are even-parity magnets, with the $l=2$ case corresponding to $d$-wave AMs. 
Odd-parity magnetic orders preserve time-reversal symmetry and therefore odd-aprity magnets cannot host a net spin polarization. 
We now examine whether geometry-induced spin polarization can arise for other even-parity magnets.

For $l=4$, corresponding to a $g$-wave AM, the altermagnetic term in the Hamiltonian is proportional to 
$k_x k_y (k_x^2-k_y^2)\si_z$. 
This term is invariant under the transformation $(k_x,k_y)\to (k_y,-k_x)$. 
As a consequence, there is no preferred momentum direction for a given spin species, and the net spin polarization vanishes in rectangular samples.

More generally, for an altermagnetic term with $l$-wave symmetry, one finds a qualitative distinction depending on whether $l=4p$ or $l=4p+2$, where $p$ is an integer. 
A typical $(4p+2)$-wave order parameter contains terms of the form
\[
[(k_x^2-k_y^2)^{2p+1-2r}(k_xk_y)^{2r}], \qquad 0\le r\le p,
\]
which are not invariant under the transformation $(k_x,k_y)\to (k_y,-k_x)$. 
The absence of this symmetry allows an imbalance between the two spin species in finite rectangular samples, leading to a net spin polarization.

In contrast, for $l=4p$ with $p$ a positive integer, the allowed terms are either of the form
\[
[(k_x^2-k_y^2)^{2p-2r}(k_xk_y)^{2r}]
\]
or
\[
[(k_x^2-k_y^2)^{2p-2r+1}(k_xk_y)^{2r-1}],
\]
both of which are invariant under $(k_x,k_y)\to (k_y,-k_x)$. 
This symmetry enforces the absence of any preferred direction for a given spin species, thereby preventing the development of a net spin polarization.

We therefore conclude that even-parity altermagnets with $(4p+2)$-wave symmetry can exhibit geometry-induced net spin polarization in rectangular samples, whereas those with $4p$-wave symmetry cannot.

\section{Summary and conclusions}

In this work, we have shown that a finite rectangular altermagnetic  sample can exhibit a net spin polarization purely due to its geometry. This effect originates from the anisotropic, spin-split bandstructure of $d$-wave altermagnets, combined with the discrete sampling of momentum space imposed by finite sample dimensions. For rectangular samples with unequal side lengths, the anisotropic discretization of $\vec{k}$ space affects the two spin species differently, leading to an imbalance in the number of occupied states and hence a finite spin polarization.

We emphasize that the spin polarization discussed here is not a boundary-localized effect that decays into the bulk. Instead, it originates from the discretization of momentum space in finite samples combined with the anisotropic spin-resolved bandstructure of altermagnets. The relevant scale governing the effect is therefore the system size itself through the momentum spacing $\Delta k_x\sim\pi/L_x$ and $\Delta k_y\sim\pi/L_y$.

We demonstrated that this geometry-induced spin polarization can be directly detected in transport measurements. In the tunneling regime, the spin polarization inferred from state counting is faithfully reflected in the charge and spin conductances which exhibit characteristic  patterns as a function of the sample dimensions. These features establish a direct correspondence between the underlying band-structure anisotropy and experimentally accessible transport observables.

Furthermore, we investigated transport across FM--AM--FM junctions and showed that the presence of a net spin polarization in the AM manifests itself as an asymmetry in the magnetoresistance under reversal of the Zeeman field in the ferromagnetic leads. This asymmetry disappears for square samples, consistent with the vanishing of the geometry-induced spin polarization when $L_x=L_y$. We also found that the asymmetry decreases with increasing system size, reflecting the fact that the spin polarization per unit area vanishes in the thermodynamic limit.

Our results highlight a route to generating and detecting spin polarization in altermagnets without invoking external magnetic fields, relying instead on sample geometry and band-structure anisotropy. This finite-size effect should be most relevant for mesoscopic devices and may offer a useful design principle for altermagnet-based spintronic applications.

\acknowledgements
The author thanks Jacob Linder and Pavlo Sukhachov for useful discussions. The author  thanks  Science and Engineering Research Board (now Anusandhan National Research Foundation) Core Research grant (CRG/2022/004311) and University of Hyderabad for financial support. 

\appendix

\section{Details of transport calculations}

In this section, we provide details of the transport calculations presented in Sec.~\ref{sec:transport}. We consider the FM--AM--FM junction described by the Hamiltonians in Eqs.~\eqref{eq:H-transport} and \eqref{eq:HFM}. In the regions $|z|>d/2$, the Hamiltonian is given by Eq.~\eqref{eq:HFM}, which reduces to that of a normal metal (NM) in the limit $b=0$. 

The problem decouples into two independent spin sectors with $\si_z=\pm 1$. In the sector $\si=\ua$ ($\si=\da$), corresponding to $\si_z=+1$ ($\si_z=-1$), the scattering eigenfunction for an electron incident from bottom to top can be written as $\psi_{\si,k_x,k_y}(z)\sin (k_x x)\sin (k_y y)$, where
\bea 
\psi_{k_x,k_y}(z) &=& 
\begin{cases}
e^{ik_{z,\si}z}+r_{\si}e^{-ik_{z,\si}z},~~{\rm for}~~z<-d/2, \\
s_{+,\si}e^{ik'_{z,\si}z}+s_{-,\si}e^{-ik'_{z,\si}z},~{\rm for}~|z|<d/2, \\
t_{\si}e^{ik_zz}, ~~~{\rm for}~~z>d/2,
\end{cases} \nn \\ && \label{eq:psi}
\eea
and $k_x=n_x\pi/L_x$, $k_y=n_y\pi/L_y$, with $n_x, n_y$ taking positive integer values. The longitudinal wave vectors are given by
$k_{z,\si}=\sqrt{(E+\si_z b)/t_na^2-k_x^2-k_y^2}$ and 
$k'_{z,\si}=\sqrt{[E-(t-t_J\si_z)k_x^2a^2-(t+t_J\si_z)k_y^2a^2]/ta^2}$. 

The scattering coefficients are determined by imposing the boundary conditions given in Eq.~\eqref{eq:bc}, together with the continuity of the wavefunction at the interfaces. The total conductance in the spin sector $\si$ is given by
\bea
 G_{\si}=(e^2/h)\sum_{n_x,n_y}|t_{\si}|^2.
\eea
Here, $t_{\si}$ depends on $n_x$ and $n_y$, and the summation runs over all positive integer values of $n_x$ and $n_y$ for which $k_{z,\si}$ is real. 

In the limit $b=0$, the charge and spin conductances are given by $G_{\ua}+G_{\da}$ and $(G_{\ua}-G_{\da})h/2e$, respectively. The resistance at Zeeman energy $b$ is defined as $R(b)=1/[G_{\ua}(b)+G_{\da}(b)]$, and the magnetoresistance is given by $MR=[R(b)-R(0)]/R(0)$.

\section{Spin polarization of a strip}

In this section, we study the spin polarization of an altermagnetic strip of finite width $L_y$ in the limit of large length $L_x$. For a fixed $L_y$, the transverse momentum is quantized, leading to a discrete set of subbands. Because the dispersion relations of the two spin sectors differ in their $k_x$ dependence, the number of subbands that intersect the Fermi energy is generally different for the two spins. As a result, the contributions of the two spin species to the total number of occupied states become unequal, giving rise to a net spin polarization.

\begin{figure}
\includegraphics[width=8cm]{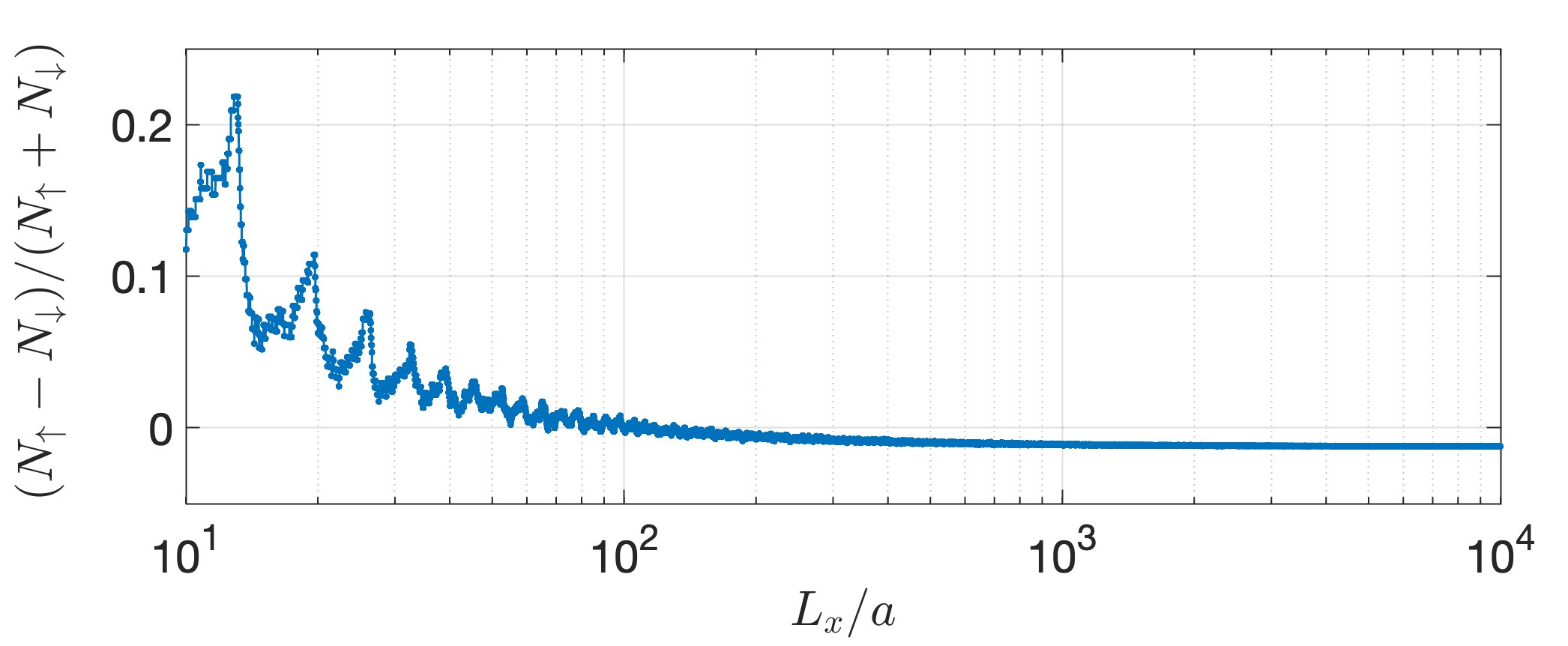}
\caption{Normalized spin polarization as a function of $L_x$ (log scale) for a fixed strip width $L_y=100a$. The spin polarization approaches a finite value as $L_x$ increases. Other parameters are the same as in Fig.~\ref{fig:Ndif}. }\label{fig:SzvsLx}
\end{figure}

In the limit $L_x\to\infty$, the momentum along the $x$ direction becomes effectively continuous, while the transverse quantization remains fixed. Consequently, the imbalance in the number of spin-resolved subbands persists, and the normalized net spin polarization approaches a finite value. Since the discretization of momentum space is identical for the two spin sectors, the difference between their populations scales with the total number of occupied states. As a result, the normalized spin polarization remains finite in the large-$L_x$ limit. This behavior is illustrated in Fig.~\ref{fig:SzvsLx}, where the spin polarization is shown to saturate to a nonzero value as $L_x$ increases for a fixed strip width $L_y$.

\bibliography{ref_ammag}

@article{das2023,
doi = {10.1088/1361-648X/acea12},
url = {https://dx.doi.org/10.1088/1361-648X/acea12},
year = {2023},
month = {aug},
publisher = {IOP Publishing},
volume = {35},
number = {43},
pages = {435302},
author = {S. Das and D. Suri and A. Soori},
title = {Transport across junctions of altermagnets with normal metals and ferromagnets},
journal = {J. Phys.: Condens. Matter}
}

@misc{sahoo2025tjde,
      title={Field-free transverse {Josephson} diode effect in altermagnets}, 
      author={Bijay Kumar Sahoo and Abhiram Soori},
      year={2025},
      eprint={2509.14109},
      archivePrefix={arXiv},
      primaryClass={cond-mat.mes-hall},
      url={https://arxiv.org/abs/2509.14109}, 
}

@ARTICLE{landauer,
  author={Landauer, R.},
  journal={IBM Journal of Research and Development}, 
  title={Spatial Variation of Currents and Fields Due to Localized Scatterers in Metallic Conduction}, 
  year={1957},
  volume={1},
  number={3},
  pages={223-231},
  doi={10.1147/rd.13.0223}
  }

@book{datta,
	author = {Datta, Supriyo },
	title = {Electronic transport in mesoscopic systems},
	publisher = {Cambridge University Press,},
	year = {2013},
	address = {Cambridge}
}

@article{linder2024,
  title = {Interface-induced magnetization in altermagnets and antiferromagnets},
  author = {Hodt, Erik Wegner and Sukhachov, Pavlo and Linder, Jacob},
  journal = {Phys. Rev. B},
  volume = {110},
  issue = {5},
  pages = {054446},
  numpages = {17},
  year = {2024},
  month = {Aug},
  publisher = {American Physical Society},
  doi = {10.1103/PhysRevB.110.054446},
  url = {https://link.aps.org/doi/10.1103/PhysRevB.110.054446}
}

@misc{weber2024,
      title={Ultrafast electron dynamics in altermagnetic materials}, 
      author={Marius Weber and Kai Leckron and Luca Haag and Rodrigo Jaeschke-Ubiergo and Libor Šmejkal and Jairo Sinova and Hans Christian Schneider},
      year={2024},
      eprint={2411.08160},
      archivePrefix={arXiv},
      primaryClass={cond-mat.mtrl-sci},
      url={https://arxiv.org/abs/2411.08160}, 
}

@article{niu2024car,
doi = {10.1088/1361-6668/ad3f56},
url = {https://doi.org/10.1088/1361-6668/ad3f56},
year = {2024},
month = {apr},
publisher = {IOP Publishing},
volume = {37},
number = {5},
pages = {055012},
author = {Ping Niu, Zhi and Mei Zhang, Yong},
title = {Electrically controlled crossed {Andreev} reflection in altermagnet/superconductor/altermagnet junctions},
journal = {Supercond. Sci. Technol.}
}

@article{ghadi2026,
doi = {10.1088/1361-648X/ae430e},
url = {https://doi.org/10.1088/1361-648X/ae430e},
year = {2026},
month = {feb},
publisher = {IOP Publishing},
volume = {38},
number = {7},
pages = {075301},
author = {Ghadigaonkar, Shubham and Das, Sachchidanand and Soori, Abhiram},
title = {Néel vector controlled charge and spin transport in altermagnetic junctions},
journal = {J.  Phys.: Condens. Matter}
}

@misc{nowak2025,
      title={Altermon: a magnetic-field-free parity protected qubit based on a narrow altermagnet {Josephson} junction}, 
      author={Sakineh Vosoughi-nia and Michał P. Nowak},
      year={2025},
      eprint={2510.18145},
      archivePrefix={arXiv},
      primaryClass={cond-mat.mes-hall},
      url={https://arxiv.org/abs/2510.18145}, 
}

@article{smejkal22b,
  title = {Beyond Conventional Ferromagnetism and Antiferromagnetism: A Phase with Nonrelativistic Spin and Crystal Rotation Symmetry},
  author = {\ifmmode \check{S}\else \v{S}\fi{}mejkal, L. and Sinova, J. and Jungwirth, T.},
  journal = {Phys. Rev. X},
  volume = {12},
  issue = {3},
  pages = {031042},
  numpages = {16},
  year = {2022},
  month = {Sep},
  publisher = {American Physical Society},
  doi = {10.1103/PhysRevX.12.031042},
  url = {https://link.aps.org/doi/10.1103/PhysRevX.12.031042}
}

@article{qfsun2024,
  title = {Field-free {Josephson} diode effect in altermagnet/normal metal/altermagnet junctions},
  author = {Cheng, Qiang and Mao, Yue and Sun, Qing-Feng},
  journal = {Phys. Rev. B},
  volume = {110},
  issue = {1},
  pages = {014518},
  numpages = {12},
  year = {2024},
  month = {Jul},
  publisher = {American Physical Society},
  doi = {10.1103/PhysRevB.110.014518},
  url = {https://link.aps.org/doi/10.1103/PhysRevB.110.014518}
}

@misc{nag2025,
      title={Field-free diode effects in one-dimensional superconductor: a complex interplay between {Fulde-Ferrell} pairing and altermagnetism}, 
      author={SVS Sai Ruthvik and Tanay Nag},
      year={2025},
      eprint={2512.01415},
      archivePrefix={arXiv},
      primaryClass={cond-mat.mes-hall},
      url={https://arxiv.org/abs/2512.01415}, 
}

@article{Das24,
  title = {Crossed {Andreev} reflection in altermagnets},
  author = {Das, Sachchidanand and Soori, Abhiram},
  journal = {Phys. Rev. B},
  volume = {109},
  issue = {24},
  pages = {245424},
  numpages = {11},
  year = {2024},
  month = {Jun},
  publisher = {American Physical Society},
  doi = {10.1103/PhysRevB.109.245424},
  url = {https://link.aps.org/doi/10.1103/PhysRevB.109.245424}
}

@article{Papaj23,
  title = {Andreev reflection at the altermagnet-superconductor interface},
  author = {Papaj, Micha\l{}},
  journal = {Phys. Rev. B},
  volume = {108},
  issue = {6},
  pages = {L060508},
  numpages = {7},
  year = {2023},
  month = {Aug},
  publisher = {American Physical Society},
  doi = {10.1103/PhysRevB.108.L060508},
  url = {https://link.aps.org/doi/10.1103/PhysRevB.108.L060508}
}

@article{sun23,
  title = {Andreev reflection in altermagnets},
  author = {Sun, Chi and Brataas, Arne and Linder, Jacob},
  journal = {Phys. Rev. B},
  volume = {108},
  issue = {5},
  pages = {054511},
  numpages = {14},
  year = {2023},
  month = {Aug},
  publisher = {American Physical Society},
  doi = {10.1103/PhysRevB.108.054511},
  url = {https://link.aps.org/doi/10.1103/PhysRevB.108.054511}
}

@article{qfsun2025,
  title = {Tunneling magnetoresistance effect in altermagnets},
  author = {Sun, Yu-Fei and Mao, Yue and Zhuang, Yu-Chen and Sun, Qing-Feng},
  journal = {Phys. Rev. B},
  volume = {112},
  issue = {9},
  pages = {094411},
  numpages = {12},
  year = {2025},
  month = {Sep},
  publisher = {American Physical Society},
  doi = {10.1103/t8b5-l859},
  url = {https://link.aps.org/doi/10.1103/t8b5-l859}
}

@article{xin2025,
  title = {Scaling behavior of magnetoresistance and Hall resistivity in the altermagnet CrSb},
  author = {Peng, Xin and Wang, Yuzhi and Zhang, Shengnan and Zhou, Yi and Sun, Yuran and Su, Yahui and Wu, Chunxiang and Zhou, Tingyu and Liu, Le and Wang, Hangdong and Yang, Jinhu and Chen, Bin and Fang, Zhong and Du, Jianhua and Jiao, Zhiwei and Wu, Quansheng and Fang, Minghu},
  journal = {Phys. Rev. B},
  volume = {111},
  issue = {14},
  pages = {144402},
  numpages = {8},
  year = {2025},
  month = {Apr},
  publisher = {American Physical Society},
  doi = {10.1103/PhysRevB.111.144402},
  url = {https://link.aps.org/doi/10.1103/PhysRevB.111.144402}
}

@article{noh2025,
  title = {Tunneling Magnetoresistance in Altermagnetic ${\mathrm{RuO}}_{2}$-Based Magnetic Tunnel Junctions},
  author = {Noh, Seunghyeon and Kim, Gye-Hyeon and Lee, Jiyeon and Jung, Hyeonjung and Seo, Uihyeon and So, Gimok and Lee, Jaebyeong and Lee, Seunghyun and Park, Miju and Yang, Seungmin and Oh, Yoon Seok and Jin, Hosub and Sohn, Changhee and Yoo, Jung-Woo},
  journal = {Phys. Rev. Lett.},
  volume = {134},
  issue = {24},
  pages = {246703},
  numpages = {6},
  year = {2025},
  month = {Jun},
  publisher = {American Physical Society},
  doi = {10.1103/nrk5-5zrj},
  url = {https://link.aps.org/doi/10.1103/nrk5-5zrj}
}
\end{document}